\documentstyle[11pt,twoside,jltp,epsfig]{article}

\def\be{\begin{equation}}
\def\ee{\end{equation}}
\def\bea{\begin{eqnarray}}
\def\eea{\end{eqnarray}}

\runninghead{M.V.Feigel'man {\it et al.}}
{Andreev Spectroscopy for Superconducting Phase Qubits}

\begin{document}

\title{Andreev Spectroscopy for Superconducting Phase Qubits}

\author{Mikhail V.\ Feigel'man 
\address{L.D.\ Landau Institute for Theoretical Physics, 
Moscow 117940, Russia}, 
Lev B.\ Ioffe 
\address{Department of Physics, Rutgers University, 
Piscataway NJ 08855, USA}, 
Vadim B.\ Geshkenbein $^{\dagger}$, and Gianni Blatter 
\address{Theoretische Physik, ETH H\"onggerberg, CH-8093 Z\"urich, 
Switzerland}}

\begin{abstract}
We propose a new method to measure the coherence time of
superconducting phase qubits based on the analysis of the
magnetic-field dependent {\it dc} nonlinear Andreev current across a
high-resistance tunnel contact between the qubit and a dirty metal
wire and derive a quantitative relation between the subgap $I$-$V$
characteristic and the internal correlation function of the qubit.


\end{abstract}

\maketitle

\section{INTRODUCTION}

While new algorithms ~\cite{Shor94,Grover97,review} are pushing the
frontier of quantum computation, their practical implementation
requires novel ideas for the design of qubits which can cope with the
contradictory constraints of scalability and long decoherence
time. Solid state electronics offers scalability but involves
macroscopic elementary blocks which usually interact strongly with the
environment and, thus, suffer from a short decoherence time. In
contrast, atomic scale designs (trapped atoms, photons in cavities,
nuclear spins) can be easily decoupled from the environment but are
difficult to combine into large and complex devices. The ideal system
for a qubit implementation is a macroscopic device with two quantum
states decoupled from the environment which can be manufactured and
combined with others by conventional techniques. This ideal is closely
approximated by the superconducting phase qubit (SCPQ) that consists
of a frustrated Josephson junction circuit \cite{Ioffe99} with two
states distinguished {\it mostly} by their superconducting phases.
\begin{figure}
\vspace{5mm}
\epsfxsize=80mm
\hspace{-00mm}
\centerline{\epsfbox{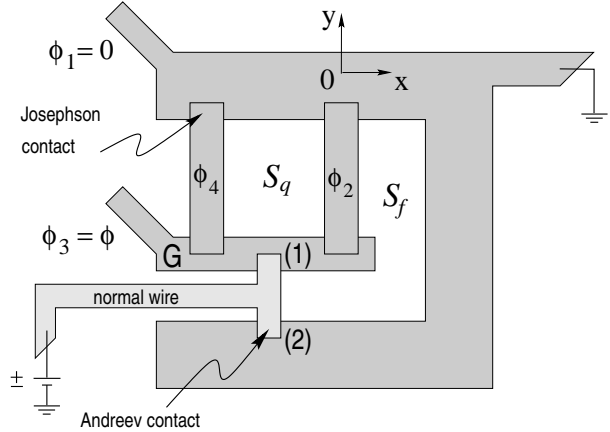}}
\vspace{-00mm}
\caption{SCPQ device with Andreev probe; the area $S_q$ of the
``intrinsic'' loop of the qubit is equal to the area $S_f$ of the
NS-QUID loop.}
\label{fig1}
\end{figure}

The simplest example of such a circuit is a {\it low inductance} SQUID
loop consisting of three (or more) Josephson junctions frustrated by a
magnetic field that creates a flux $\Phi =\Phi_0/2$ through the
loop\cite{Mooij99}, see Fig.\ 1. The two states of the qubit are
defined by the phase changing clockwise or counterclockwise around the
loop. The advantage of an ``all phase" design is that the
superconducting phase is not directly coupled to the classical
environment; furthermore, superconductors themselves have a gap such
that the environment has a low density of low energy modes potentially
leading to decoherence and the charge fluctuations are screened such
that the electric coupling to the environment is negligible. The
magnetic coupling to the environment cannot be completely avoided as
the phase changing around the frustrated loop is coupled to an induced
current, however, the latter can be made small if the Josephson
junctions are weak and the loop inductance is small.  In particular,
we shall show that the interaction of these currents with the
environment leads to a small decoherence even if the environment
includes potentially dangerous normal wires. Finally, we note that the
coupling with the frustrating magnetic field present in the simplest
design can be eliminated in designs involving $d$-wave
superconductors\cite{Ioffe99} or $\pi-$junctions\cite{Geshkenbein99}.

The first problem to be addressed in the development of a SCPQ is the
implementation of a convenient probe which tests whether the device
produces coherent Rabi oscillations. It is obviously difficult to test
a system which is decoupled from the environment and it is even more
difficult to check that it remains in the phase coherent state without
disturbing it. Here we suggest an experiment to probe the state of
such qubits and to determine their decoherence time. The idea is to
measure the Andreev conductance between the ground and a normal wire
in the geometry of a fork with the two prongs connecting to two
separate islands of the qubit (Fig.\ 1). If the qubit phase is fixed
(classical regime), the low-$T$ conductance exhibits a contribution
from processes in which electrons from the normal metal are reflected
as holes from the qubit boundary, diffuse to the bulk superconductor
boundary and are reflected from it as electrons. Such a phase coherent
electron diffusion in the normal metal prongs of the ``fork'' leads to
periodic conductance oscillations (with period $\Phi_0$) as a function
of the magnetic flux penetrating through the $S_f$ region of the
fork.\cite{Nazarov93,Pothier94} This is due to the fact that the
magnetic field controls the superconducting phase difference between
the two NS contacts of the fork [(1) and (2) in Fig.\ 1] and thus
influences the electron interference picture.  A similar mechanism
controls the experiment discussed in the quantum regime when the phase
of the qubit fluctuates.  Qualitatively, at small bias voltage the
quantum phase acquired by the reflected electrons fluctuates and the
resulting contribution to the conductivity averages out. At larger
voltages the phase fluctuations become slow compared with the electron
tunneling time and the contribution to the conductivity is
restored. Below we describe the specific device parameters and derive
quantitative formulas describing this effect. A similar proposal ---
to probe the Josephson splitting of levels in a superconducting single
Cooper-pair transistor by measuring the Andreev conductance --- was
put forward in Ref.\ \citen{Glazman95}.

\section{SUPERCONDUCTING PHASE QUBIT AND ENVIRONMENT}

Consider the SQUID loop with four junctions shown in Fig.\ 1. Two
degenerate states naturally appear in such a loop if the flux $\Phi_q$
of the external field through the qubit loop is exactly $\Phi
_0/2=hc/4e$.  Indeed, in a gauge with $A_x=0$, $A_y=Bx$ the classical
minima of the Josephson energy correspond to the phase drops $0$ or
$\pi$ around the loop; in these states the phase of the island $G$
(see Fig.\ 1) is $\phi^{\pm}=\pm \pi /2$ (relative to the phase at
point $0$); below we refer to these states as $|\!\uparrow\rangle$ and
$|\!\downarrow\rangle$. In order to reduce the parasitic coupling to
the environment, the inductance $\cal L$ of the loop should be
sufficiently small so that ${\cal L}I_c/c\leq 10^{-3}\Phi_0$, $I_c$ is
the critical current.  In the absence of any charging energy in the
junctions the system prepared in one of these states will stay put
forever; quantum effects are due to the small but finite charging
energy $E_C=e^2/2C$ determined by the capacitances $C$ of the
junctions, which should be smaller than the Josephson energy $E_J
\equiv \hbar I_c/2e \gg E_C$.  The tunnelling frequency between the
two classically degenerate ground states is estimated as $\Omega_0
\approx \omega_0\exp(-a\sqrt{\hbar I_cC/e^3})$, where
$\hbar\omega_0=\sqrt{8E_CE_J}$ is the Josephson plasma frequency of
the contacts; the coefficient $a \approx 1.6$ was found by a numerical
evaluation of the saddle-point trajectory in the space of the three
phase variables $\phi_2$, $\phi_3 \equiv \phi$, and $\phi_4$. We
assume values of $I_c$ in the range $10^{-7}$ A, and capacitances of
order of a few fF, resulting in $\omega_0/ 2\pi\sim 100$ GHz and a
tunneling frequency $\Omega_0/2\pi$ in the range of few GHz. Our
choice of parameters is limited by the following constraints: (a) the
system should be sufficiently small to fulfill the condition ${\cal
L}I_c/c \leq 10^{-3}$, but not too small, since (b) we need $E_C \ll
E_J$; finally, (c) $\Omega_0$ should not be too small, in order to
keep the decoherence time $t_{\rm dc}$ much longer than
$\Omega_0^{-1}$.

Once tunnelling is taken into account, the true eigenstates become
$|\,0\rangle=[|\!\uparrow\rangle+|\!\downarrow\rangle]/\sqrt{2}$ and
$|\,1\rangle=[|\!\uparrow\rangle-|\!\downarrow\rangle]/\sqrt{2}$ and
are separated by the energy gap $\hbar\Omega_0$ (here and below we
shall use the ``spin" notation to describe the {\it classically}
degenerate states). The deviation of the magnetic flux $\Phi_q = B
S_q$ through the loop from the point $\Phi_q=\Phi_0/2$ removes the
degeneracy of the states $|\!\uparrow\rangle$ and
$|\!\downarrow\rangle$. Thus the intrinsic Hamiltonian of our qubit
written in the basis $|\!\uparrow\rangle,|\!\downarrow\rangle$ is
$H=h_x\sigma_x+ h_z\sigma_z$, where
$h_z=(I_c/\sqrt{2}c)[\Phi_q-\Phi_0/2]$ and $h_x=\hbar\Omega_0$. All
operations on the qubit can be performed if one is able to vary the
effective fields $h_z$ and $h_x$.  The variation of $h_z$ can be
achieved by changing the flux $\Phi_q$ in the loop, while the
variation of $h_x$ can be implemented by two means: the variation of
the gate potential applied to the island $G$ changes this field
smoothly (c.f., Ref.\ \citen{Ivanov98}), while switching a capacitor
$C_{\rm ext}\sim 10C$ in parallel blocks the tunneling abruptly.

The coupling of a two-level quantum system to the environment is
usually described by the Caldeira-Leggett ``spin-boson" model where
the ``spin" (i.e., two-level system, TLS) is linearly coupled to an
ensemble of oscillators.\cite{Chak84,RMP85,Weiss98} Actually, our TLS
originates from the dynamics of a continuous phase variable $\phi$
with a potential energy $U(\phi)$ strongly favoring values $\phi$
around $\pm\phi_0/2$. Then the term that describes the coupling to the
environment can be written, upon integration over environmental modes,
in the form of a non-local imaginary-time action $S_{\rm diss} =
\frac12 \int dt \, dt' K(t-t') [\phi(t)-\phi(t')]^2$.  While in a
superconducting system we can hope the low energy spectrum of the
environment to be gapped providing favorable conditions for a long
coherence time, here, we analyze the (worst case) situation where the
coupling of low lying modes survives to produce a finite ohmic
dissipation with $K(t) = \eta/2\pi t^2$, where $\eta$ is the
generalized ``friction coefficient" in the equation of motion $\eta
\dot\phi = -\partial U(\phi)/\partial\phi$.  The strength of the
environmental coupling is determined by the dimensionless parameter
$\alpha = \eta\phi_0^2/2\pi\hbar$;
consider the symmetric system with $h_z = 0$: for $\alpha > 1$ the
tunnelling between the states $|\!\uparrow\rangle$, and
$|\!\downarrow\rangle$ is suppressed and the system possesses two
classically degenerate ground states. For intermediate coupling $1/2 <
\alpha < 1$ tunnelling is finite but {\it incoherent}: at time scales
$t > \Omega^{-1} = \hbar/h_x$ (with $\Omega$ the effective tunneling
frequency depending on $\alpha$) the TLS is described by the classical
master equation for the occupation probabilities
$P_{\uparrow,\downarrow}$ of the two states with a transition rate
$\Gamma(\alpha)$ vanishing at $\alpha \to 1$. At $h_z=0$ this equation
takes the form $\dot{m}(t) = - \Gamma m$ and the ``magnetization"
$m(t)=\langle\sigma_z\rangle(t) = P_\uparrow(t) - P_\downarrow(t)$
monotonically decays to zero.  Here, we are interested in the case of
very weak coupling $\alpha \ll 1$, where the TLS dynamics is weakly
perturbed by the environment~\cite{RMP85}, i.e., the decoherence time
$t_{\rm dc} = \Gamma^{-1}$ shall be much longer than the period of
coherent oscillations $2\pi\Omega^{-1}$. In this case, the
magnetization exhibits damped oscillations $m(t) = \exp(-\Gamma
t)\cos(\Omega t)$ with a damping rate $\Gamma$ and a frequency
$\Omega$ determined by the parameter $\alpha < 1/2$
via~\cite{Chak84,Lesage98}
\begin{equation}
\Omega = (\Omega_r/\pi \alpha) \sin[\pi\alpha/(1-\alpha)], 
\qquad
\Gamma = \Omega/Q = \Omega \tan[\pi\alpha/2(1-\alpha)],
\label{lesage}
\end{equation}
with the renormalized tunnelling frequency $\Omega_r = \Omega_0
(\Omega_0/\omega_c)^{\alpha/1-\alpha}$; here, $\omega_c$ is an upper
cutoff for the frequencies of the environmental modes.  A measurement
of $m(t)$, particularly its quality factor $Q$, then allows us to test
the degree of quantum coherence in the device.  

The direct determination of $m(t)$ in a time-domain experiment like
the one reported in Ref.\ \cite{Nakamura99} (where a superconducting
``single Cooper-pair" device was studied) is a demanding task. As an
alternative one may attempt to find the correlation function $D(t)=
\langle\sigma_z(t)\sigma_z(0)\rangle$ through a
measurement\cite{Nakamura98} of the {\it rf} absorption; indeed, the
latter provides the imaginary part $\chi^{{\prime\prime}}(\omega)$ of
the response function $\chi(\omega)=2\int_0^{\infty} dt
D^a(t)\exp(i\omega t)$, where $D^a(t) =
\langle[\sigma_z(0),\sigma_z(t)] \rangle/2i$ is the antisymmetric part
of the spin-spin correlation function.  Standard relations then allow
us to extract the correlator $D(t)$ from the measured absorption
$\chi^{{\prime\prime}}(\omega)$: the symmetrized correlation function
$D^s(t)=\langle \{\sigma_z(0), \sigma_z(t)\}\rangle/2 =
\coth(\hbar\omega/2T)\,\chi^{{\prime\prime}} (\omega)$ relates to the
correlator via $D(\omega)=2D^s(\omega)/ [1+\exp(-\omega/T)]$ (note
that $D(\omega < 0) = 0$ at $T=0$).  Below we discuss yet another
experiment giving us access to $D(t)$: we will see that the nonlinear
Andreev conductance $dI_{\rm\scriptscriptstyle A}/dV$ through the
qubit's island $G$ can be expressed via the Fourier-transform
$D(\omega)$ of its correlator. Scanning magnetic field and voltage in
a transport measurement then appears to give a much easier and direct
access to the operation of the qubit than the more usual real-time and
{\it rf}-techniques.

Being not aware of exact results available for $\chi(\omega)$ or
$D^s(t) = \langle \{\sigma_z(0),\sigma_z(t)\}\rangle/2$, we make use
of an approximate analysis~\cite{Volker98} to arrive at an idea of the
relevant dependencies in these quantities: for $Q \gg 1$ the
absorption characteristic is close to a combination of Lorentzians:
$\chi^{{\prime \prime}}(\omega) = \Gamma/[\Gamma^2 +
(\omega-\Omega)^2] -\Gamma/[\Gamma^2 + (\omega + \Omega)^2]$, implying
a symmetrized correlator of the form $D^s(\omega) =
4|\omega|\Omega\Gamma/ [(\Gamma^2+\Omega^2 +\omega^2)^2 -
4\omega^2\Omega^2]$.  In the time domain, $D^s(t)|_{T=0} \approx
\cos(\Omega t)\,\exp(-\Gamma t) + 4/[\pi Q(\Omega t)^2]$, where the
last term is due to the non-analyticity of $D^s(\omega)$ at
$\omega=0$.

\section{ANDREEV CONDUCTANCE}

We first concentrate on the subgap conductance through a NIS contact
(\#1 in Fig.\ 1) with a low normal-state conductance $\sigma_t = G_t
e^2/\hbar \ll e^2/\hbar$ and discuss the effect of fluctuations in the
phase $\phi(t)$ on the island $G$. Consider an ideal SCPQ operating at
the degeneracy point ($h_z = 0$) with the ground and excited states
$|\,\!0\rangle$ and $|\,\!1\rangle$ separated in energy by
$\hbar\Omega$. The operator $C^\dagger = \exp(-i\phi)$ creating Cooper
pairs changes the relative sign of the states $|\!\uparrow \rangle$
and $|\!\downarrow \rangle$, thus $C^\dagger|\,\!0\rangle =
|\,\!1\rangle$ and $C^\dagger$ has no matrix element in the ground
state. Therefore $|\,\!0\rangle$ is a state with a well-defined number
of Cooper-pairs {\it modulo} 2, implying that the usual Andreev
process transmitting a single Cooper pair through the interface is
forbidden at low voltage $V < \hbar\Omega/2e$, while at voltages above
the gap $\hbar\Omega/2$ the Andreev conductance will be
large. Higher-order processes with simultaneous tunnelling of two
Cooper pairs are allowed, but their rate is of the order of $G_t^4$
and thus is negligible at $G_t \ll 1$. Overall, we then expect an
Andreev conductance $dI_{\rm\scriptscriptstyle A}/dV$ similar to the
one of a usual NIS junction but with the superconducting gap $\Delta$
replaced by the two-level separation $\hbar\Omega$. On the other hand,
for a fixed phase on the island we can expect to observe the usual
Andreev current across a NIS junction.

The above consideration defines the simplest example of a situation
where the quantum fluctuations of the superconducting phase suppress
the Andreev conductance of the SIN contact. General expressions
describing these effects have been derived in Ref.\
\citen{Feigelman99} (c.f., Appendix B). Here, we discuss the result
for the nonlinear subgap current $I_{\rm\scriptscriptstyle A}(V)$ for
the case where the normal metal is relatively clean (with a
dimensionless conductance $g \gg 1$) such that all corrections of
relative order $1/g$ can be neglected,
\begin{eqnarray}
I_{\rm\scriptscriptstyle A}(V) &=& 
\frac{e G_t^2}{4\hbar}\int_{-\infty}^{\infty} C(E) dE
  \int_{-\infty}^{\infty}\frac{dE'}{2\pi\hbar} D_\phi(E') 
\frac{1-e^{-2eV/T}}{1-e^{(E'-2eV)/T}} \label{main}\\ 
&&\times
\left[\tanh[(E-E'/2+eV)/2T]-\tanh[(E+E'/2-eV)/2T]\right].
\nonumber
\end{eqnarray}
Here, $C(E)$ is the real part of the Cooperon amplitude in the normal
metal (see below) and $D_\phi(E) = \int dt\, \exp(iEt/\hbar)
D_\phi(t)$ is the Fourier-transformed autocorrelation function
$D_\phi(t) = \langle \Psi(t)\Psi^+(0) \rangle$ of the superconducting
order parameter $\Psi(t) = \exp[i\phi(t)]$ on the island $G$. On long
time scales $t \gg \Omega^{-1}$ the oscillatory modes near the minima
of the Josephson potential are irrelevant; the values of $\phi$ in
these minima differ by $\pi$, thus $D_\phi(t)$ is equivalent to the
(sought for) TLS correlation function $D(t)$ discussed in the previous
section.

The real part of the Cooperon amplitude is given by
\begin{equation}
\label{Cooperb}
C(E) = \frac{1}{\nu V}  \Re \sum_{q}\frac{1}{Dq^2 - 2i E/\hbar
+ \tau_{\varphi}^{-1}},
\end{equation}
where the sum goes over the Cooperon eigenmodes $q_n$, $V$ denotes the
volume, $\nu$ the density of states (for a single projection of spin),
and $D$ the diffusion constant of the normal metal.  Assuming a wire
geometry with $w$, $L \gg (\min(L_E \equiv \sqrt{\hbar D/4E},
L_\varphi \equiv \sqrt{D\tau_\varphi}),w)$, and $d \leq w$ denoting
its length, width and thickness, Eq.\ (\ref{Cooperb}) reduces to the
expression $C(E)\approx \rho L_E/[1+(\hbar/4
E\tau_\varphi)^2)]^{1/4}$, where the resistance $\rho=(2\nu D
wd)^{-1}$ per unit length of the wire is measured in units of
$(\hbar/e^2)/ {\rm Length}$. The term $\propto \tau_\varphi^{-1}$
describes the electron dephasing present at all $T > 0$ due to
electron-electron interactions and external
noise~\cite{Gershenzon99}. For a short wire, $C(E) = \rho L$; in the
end we arrive at the simple result $C(E) \approx \rho
\min(L_E,L_\varphi,L)$. An expression for $I_{\rm\scriptscriptstyle
A}(V)$ analogous to Eqs.\ (\ref{main}) has been derived
earlier~\cite{Huck98} for a similar physical problem.

Let us analyze the result (\ref{main}) in the limit $T\to 0$ which
applies to the regime $T \ll (eV,\hbar\Omega)$. At the same time we
keep the electron dephasing rate $\tau_{\varphi}^{-1}$ nonzero, since
experimentally~\cite{Gershenzon99} an apparent saturation at $T \leq
0.1-1 K$ has been found. Differentiating Eq.~(\ref{main}) by $V$ at
$T=0$ we find
\begin{equation}
\frac{dI_{\rm\scriptscriptstyle A}}{dV} = 
\frac{e^2G_t^2}{2\pi\hbar^2}
\int_{0}^{2eV}{dE}\, D_\phi(E)\, C(eV-E/2)
\label{zeroT}
\end{equation}
(values $E<0$ are excluded from the integral since 
$D_\phi(E<0)=0$~at~$T=0$).

Consider first of all the degenerate case with $h_z=0$ in the absence
of any dephasing within the qubit, i.e., $\Gamma = 0$ and $D_\phi(E) =
2\pi\delta(E/\hbar-\Omega)$. The evaluation of the integral in
(\ref{zeroT}) is trivial and we obtain the result
\begin{equation}
\frac{dI_{\rm\scriptscriptstyle A}}{dV} = \cases{\qquad\qquad 0, 
& $eV < \hbar\Omega/2$,\cr
(e^2/\hbar)\,G_t^2\,C(eV-\hbar\Omega/2), 
& $\hbar\Omega/2 < eV$,}
\label{dIdVzeroT}
\end{equation}
in agreement with the above qualitative arguments. In the following we
restrict the discussion to long wires $L \gg \sqrt{D/\Omega}$; then
the conductivity above threshold shows a square-root singularity which
is smeared by a nonzero dephasing rate $\tau_\varphi^{-1}$.  The
nonzero conductance below threshold is due either to a finite
dissipation rate $\Gamma > 0$ or a finite temperature. At zero
temperature, a finite $\Gamma$ produces the sub-threshold ($eV <
\hbar\Omega/2$) conductance
\[
\frac{dI_{\rm\scriptscriptstyle A}}{dV} =
\frac{e^2}{\hbar}G_t^2 \rho \sqrt{\frac{D}{\Gamma}} 
\cases{
\displaystyle{
a \left(\frac{\sqrt{\hbar\Gamma\,eV}}
{\hbar\Omega}\right)^3
             }, & $eV \ll \hbar\Omega/2$,\cr 
\displaystyle{
\frac{1}{2}\left[\frac{\hbar\Gamma}{\delta}
\frac{\sqrt{1+\frac{\hbar^2\Gamma^2}{\delta^2}}-1}
{1+\frac{\hbar^2\Gamma^2}{\delta^2}}\right]^{1/2},
             }
& $\delta \equiv \hbar\Omega - 2eV \ll \hbar\Omega/2$,
}
\]
where we have assumed $\tau_\varphi \rightarrow \infty$ and have made
use of the form of $D_\phi(E)$ discussed at the end of Sec.~2. The
numerical coefficient $a$ depends on the shape of the correlator
$D_\phi(E)$ and is of order unity; for the Lorentzian shape discussed
above we find $a = 32/3\pi$.  A finite phase breaking time
$\tau_\varphi$ in the normal wire changes this result at low voltages
$eV < \hbar/\tau_\varphi$: $dI_{\rm\scriptscriptstyle A}/dV \approx
(e^2/\hbar) G_t^2\rho L_\varphi [\hbar\Gamma (eV)^2/
(\hbar\Omega)^3]$. Finally, a finite temperature also produces a
finite sub-threshold conductance which, however, can be well separated
(at $ T \ll (2eV, \hbar\Omega - 2eV)$) from the above behavior due to
its exponential $\propto \exp[(2eV-\hbar\Omega)/T]$ rather than
algebraic dependence $\propto (\sqrt{\hbar\Gamma\, eV}/
\hbar\Omega)^{3}$ and $\propto \hbar\Gamma/ (\hbar\Omega-2eV)$ at low
and high voltages.  Note, that these results for the suppression in
the conductance $dI_{\rm\scriptscriptstyle A}/dV$ due to phase
fluctuations refer to terms of order $G_t^2$ and hence our analysis is
valid only for low $G_t$.

The above discussion clarifies the appearance of a threshold behavior
in the degenerate case $h_z = 0$, where the differential conductance
$dI_{\rm\scriptscriptstyle A}/dV$ sharply drops below the gap
$\hbar\Omega/2$ and vanishes at $V,~T \to 0$. For $h_z \geq h_x =
\hbar\Omega$ the correct ground state is close to a state with a
well-defined phase $\phi$ and the conductance attains its
semiclassical value $dI_{\rm\scriptscriptstyle A}^{\rm cl}/dV =
(e^2/2\hbar)G_t^2\rho \sqrt{\hbar D/eV}$. As a consequence, a sharp
minimum (at $\Phi_q=\Phi_0/2$ corresponding to $h_z=0$) is expected in
$I_{\rm\scriptscriptstyle A}(V)$ when changing the magnetic flux
$\Phi_q$ through the qubit. This qualitative conclusion survives the
presence of a finite electron decoherence rate $\tau_\varphi^{-1}$ in
the normal-metal, however, the precise form of
$dI_{\rm\scriptscriptstyle A}/dV$ is modified. A short electron
decoherence time $\tau_\varphi \leq \hbar/T$ is frequently observed in
thin wires at lowest temperatures as relevant for the experiments on
qubits (c.f., Ref.\ \citen{Gershenzon99}).  These observations are
poorly understood theoretically and thus it is more practical to
measure $\tau_\varphi$ independently in the same experimental
setup. Such a measurement can be carried out via an analysis of the
magnetic field dependent Andreev conductance in the ``fork" geometry
as we are now going to discuss.

Consider the interference experiment for the device shown in Fig.\
1. The first stage of the proposed experiment is to measure the
amplitude $I_{\rm\scriptscriptstyle A}^{\rm\scriptscriptstyle (12)}$
in the oscillations of the Andreev current $I_{\rm\scriptscriptstyle
A} = I_{\rm\scriptscriptstyle A}^{\rm\scriptscriptstyle (1)} +
I_{\rm\scriptscriptstyle A}^{\rm\scriptscriptstyle (2)} +
I_{\rm\scriptscriptstyle A}^{\rm\scriptscriptstyle (12)}$ as a
function of magnetic field $B$ with the aim to extract the electron
decoherence time $\tau_\varphi$ in the normal metal (the superscripts
$^{(i)}$ refer to the contributions from the contacts $i=1,2$, see
Fig.\ 1). The total phase determining the interference current
$I_{\rm\scriptscriptstyle A}^{\rm\scriptscriptstyle (12)}$ sums up to
$\phi_{\rm\scriptscriptstyle A} = \phi + 2\pi BS_f/\Phi_0$. Away from
the degeneracy field $B_n \equiv (n+1/2)\, \Phi_0/S_q$, the
(semiclassical) phase $\phi$ in our device is determined by the
minimization of the Josephson energy and we obtain $\phi=\pi(\{B
S_q/\Phi_0+1/2\}-1/2)$, where $\{x\}$ is the fractional part of $x$.
Choosing a geometry with $S_q=S_f$, we find for $B S_q/\Phi_0 \in
[n,n+1/2]$ an interference current $I_{\rm\scriptscriptstyle
A}^{\rm\scriptscriptstyle (12)}=2j_{\rm\scriptscriptstyle
12}\cos\phi_{\rm\scriptscriptstyle A} = 2j_{\rm\scriptscriptstyle
12}\cos(3\pi B S_q/\Phi_0)$, whereas for $B S_q/\Phi_0 \in
[n+1/2,n+1]$ the sign changes: $I_{\rm\scriptscriptstyle
A}^{\rm\scriptscriptstyle (12)}=-2j_{\rm\scriptscriptstyle
12}\cos(3\pi B S_q/\Phi_0)$. The amplitude~\cite{Nazarov93} of the
interference component $j_{\rm\scriptscriptstyle 12}$ depends on the
precise geometry of the ``fork" and on the electron dephasing length
$L_\varphi$ which should be longer than the distance
$L_{\rm\scriptscriptstyle 12}$ between the contacts \#1,2 (otherwise
$j_{\rm\scriptscriptstyle 12}$ will be exponentially small).  Detailed
calculations and measurements of $j_{\rm\scriptscriptstyle 12}$ have
been presented by the Saclay group~\cite{Pothier94,Pothier294,Gueron};
they found a discrepancy by a factor 1.8 between the best fit (with
$\tau_\varphi$ as fitting parameter) to their data and the calculated
expression for $j_{\rm\scriptscriptstyle 12}$ (c.f., Ref.\
\citen{Gueron}, p.\ 190; this discrepancy may be due to the use of a
simplified description of the electron decoherence using a factor
$\exp(-t/\tau_\varphi)$ in the real time expression for the
Cooperon~\cite{Pothier94,Pothier294}).
\begin{figure}
\vspace{-00mm}
\epsfxsize=65mm
\hspace{-00mm}
\centerline{\epsfbox{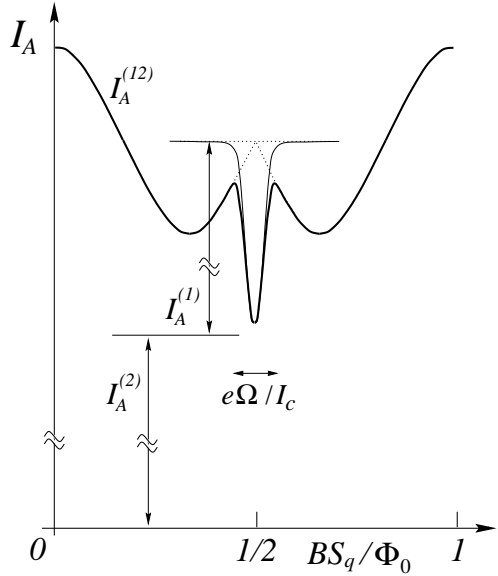}}
\vspace{-00mm}
\caption{Schematic view of the Andreev current through the ``fork"
enclosing the qubit versus magnetic flux for the special case
$S_q=S_f$.  The fork serves two purposes: the interference pattern
gives access to $\tau_\varphi$ through measuring
$I^{\rm\scriptscriptstyle (12)}_{\rm\scriptscriptstyle A}$ and allows
for the determination of the exact point of degeneracy.}
\label{fig2}
\end{figure}
Second, we test for the coherent time evolution of the qubit: In the
geometry with $S_q=S_f$, the semiclassical current amplitude
$I_{\rm\scriptscriptstyle A}^{\rm\scriptscriptstyle (12)}(B)$ is
periodic with a period $\Phi_0/S_q$, showing upward cusps at $B =
B_n$, see Fig.\ 2, dotted line. These cusps are due to the penetration
of single flux quanta $\Phi_0$ into the qubit loop at $B=B_n$. At
these fields, the interference contribution $I_{\rm\scriptscriptstyle
A}^{\rm\scriptscriptstyle (12)}$ vanishes and the semiclassical
Andreev current is given by the sum $I_{\rm\scriptscriptstyle
A}^{\rm\scriptscriptstyle (1)} + I_{\rm\scriptscriptstyle
A}^{\rm\scriptscriptstyle (2)}$ (in the generic case $S_q\neq S_f$ the
component $I_{\rm\scriptscriptstyle A}^{\rm\scriptscriptstyle (12)}$
does not vanish at $B=B_n$ and the flux penetration leads to a jump in
the measured $I_{\rm\scriptscriptstyle A}$). In addition, at $B = B_n$
the quantum fluctuations in the phase $\phi$ become large and suppress
the Andreev current $I_{\rm\scriptscriptstyle
A}^{\rm\scriptscriptstyle (1)}$ to the ``active" island G of the SCPQ,
see Eq.\ (\ref{zeroT}).  Hence, close to the points $B=B_n$ a narrow
dip is superimposed upon the above-mentioned cusp in the
$I_{\rm\scriptscriptstyle A}(B)$ dependence, see Fig.\ 2, solid
line. The relative width $\Delta B/B_0 \sim e\Omega/I_c$ of this dip
is determined by the relation $h_z=I_c (\Phi_q-\Phi_0/2)/ \sqrt{2}c
\leq h_x=\hbar\Omega$ and provides the coherence gap of the qubit.  In
addition, we can measure the decoherence rate $\Gamma$ via an analysis
of the conductance $dI_{\rm\scriptscriptstyle
A}/dV=\sigma_{\rm\scriptscriptstyle A}(V)$ at the degeneracy points
$B=B_n$.~In the limit $T \ll eV$ and $eV \to 0$ the entire conductance
is determined~by~the current through the contact \#2,
$\sigma_{\rm\scriptscriptstyle A}^{\rm\scriptscriptstyle (2)}(V) =
(e^2/\hbar) G_t^2 C(eV)$.  The variation of the full differential
conductance $\sigma_{\rm\scriptscriptstyle A}(V) =
\sigma^{\rm\scriptscriptstyle (1)}_{\rm\scriptscriptstyle A}(V) +
\sigma^{\rm\scriptscriptstyle (2)}_{\rm\scriptscriptstyle A}(V)$ at
low voltages $eV \leq \hbar\Omega$ then can be used to extract the
value of $\Gamma$. Note that $C(E)$ for the fork differs from the form
used in the expression for the subgap conductance, c.f., Ref.\
\citen{Pothier294}.  The detailed analysis of
$\sigma_{\rm\scriptscriptstyle A}(V)$ for the different relations
between $\Omega$, $\Gamma$, and $\tau_\varphi$ will appear in a future
publication.

Finally, any method measuring the SCPQ coherence time is useful only
if the presence of the measurement circuit does not, by itself, lead
to a decoherence rate comparable to the ``intrinsic" one. In our setup
there are two sources of additional phase $\phi$ decoherence due to
the presence of the normal-metal wire nearby: (i) the direct coupling
of the qubit to the environment via the non-zero Andreev conductance,
and (ii) the inductive coupling between {\it ac} supercurrents in the
qubit loop and normal currents in the Andreev fork. We consider them
one by one: (i) Comparing the dissipative action for the Andreev
conductance (c.f., Ref.\ \citen{FL} for example) with the general
relation for $S_{\rm diss}$ and the definition of the parameter
$\alpha = \eta\phi_0^2/2\pi\hbar$ one finds the contribution
$\alpha_{\rm\scriptscriptstyle A} =
(\pi\hbar/8e^2)\sigma_{\rm\scriptscriptstyle A}^{\rm cl}$, where
$\sigma_{\rm\scriptscriptstyle A}^{\rm cl} = (e^2/\hbar) G_t^2 \rho
\min(L,L_\varphi)$ is the semiclassical Andreev conductance via the
contact \#1.  Thus, a classical subgap resistance of the SN contacts
in the M$\Omega$ range will produce a value
$\alpha_{\rm\scriptscriptstyle A} \leq 10^{-3}$.  (ii) We calculate
the friction coefficient $\eta_{\rm ind}$ via the energy dissipation
rate $W_{\rm ind} = \eta_{\rm ind} (d\phi/dt)^2$. The main
contribution to $W_{\rm ind}$ is due to the {\it ac} magnetic field
applied to the conducting wire (with thickness $d < 50$ nm, width $w
\sim 100$ nm and a few microns length $L$), which gives $W_{\rm ind}
\approx \omega (wLd)\chi^{\prime\prime}(\omega)B_\omega^2$, where
$\chi^{\prime\prime}(\omega) \sim \omega w^2/c^2$ is the imaginary
part of the magnetic susceptibility of a thin-film wire in a magnetic
field $B_\omega$ in the low-frequency limit (for frequencies $\omega
\sim 10^{10}$ and a conductivity $\sigma \sim 10^{-5}~\Omega$cm of the
wire the effective skin depth is much longer than the wire's width $w
\sim 100$ nm). The magnetic field outside the SCPQ loop (of size
$L_q\sim 1\mu$m) is less than $\Phi_q/L_q^2$, thus $(\omega
B_\omega)^2 < (\omega\phi_\omega)^2 (\hbar/e)^2 (\Phi_q/
\Phi_0)^2L_q^{-4}$. Combining all factors, one finds $\alpha_{\rm ind}
< (w^2L/\rho L_q^4)(\Phi_q/\Phi_0)^2$. For a SCPQ with $\Phi_q/\Phi_0
< 10^{-3}$ and $w/L_q \sim 0.1$ this results in a negligible value
$\alpha_{\rm ind} < 10^{-7}$.

In conclusion, we have derived expressions for the subgap $I(V)$
characteristics, Eqs.\ (\ref{main}) and (\ref{zeroT}), which can be
used for the determination of the intrinsic correlation function
$D_\phi = \langle \exp[i\phi(t)]\,\exp[-i\phi(0)]\rangle$ of a
superconducting phase qubit (SCPQ), once the electron dephasing time
$\tau_\varphi$ that enters the Cooperon, Eq.\ (\ref{Cooperb}), is
known. As the latter is difficult to control theoretically at the
ultralow temperatures of the experiment we propose an interference
experiment probing the SCPQ with an ``Andreev fork", allowing to
determine $\tau_\varphi$ within the same setup and then use its value
to extract $t_{\rm dc}$.  In the absence of decoherence in the SCPQ,
$\Gamma=0$, the current $I_{\rm\scriptscriptstyle A}(V)$ is zero (at
$T=0$) below the threshold $eV=\hbar\Omega/2$, irrespective of the
value of $\tau_\varphi$.  Thus the SCPQ decoherence time $t_{\rm dc} =
\Gamma^{-1}$ can be measured even if it is longer than $\tau_\varphi$.
In practice, the most important limitation of this technique seems to
be due to the non-vanishing temperature of experiment which produces a
nonzero sub-threshold conductance even at $\Gamma=0$.  However, in the
range $T \ll \hbar\Omega$ and at voltages such that $T \ll (2eV,
\hbar\Omega-2eV)$ the behavior of $dI_{\rm\scriptscriptstyle A}/dV$ is
exponentially close to the zero-$T$ one; therefore even values of
$\Gamma \ll T/\hbar$ can be extracted from the measured
$dI_{\rm\scriptscriptstyle A}/dV$ curves.

We thank G.\ Lesovik, J.\ Mooij, M.\ Skvortsov, and C.\ van der Wal 
for helpful discussions and the `Zentrum f\"ur theoretische Studien' 
at ETH-Z\"urich for hospitality and financial support. 
The research of M.V.F.\ was supported by the RFBR grant 98-02-19252 
and by the Russian Ministry of Science via the programm 
``Physics of Quantum Computing".

\end{document}